\documentstyle[aps]{revtex}

\begin{document}
\flushbottom
\def\thepage{\roman{page}}
\title{\vspace*{1.5in} Non-linear metric perturbations and production
of primordial black holes}
\author{P. Ivanov}

\address{Theoretical Astrophysics Center, Juliane Maries Vej
30, 2100 Copenhagen, {\O} Denmark\\
Astro Space Center of P. N. Lebedev Institute, Profsoyznaya 84/32,
117810 Moscow, Russia}
\maketitle

\input{epsf}

\newcommand{\be}{\begin{equation}}
\newcommand{\ee}{\end{equation}} \newcommand{\g}{\nabla}
\newcommand{\de}{\partial}    \newcommand{\ha}{\frac{1}{2}}
\newcommand{\ci}[1]{\cite{#1}}  \newcommand{\bi}[1]{\bibitem{#1}}
\newcommand{\noi}{\noindent}

\newcommand{\ga}{\alpha}
\newcommand{\gb}{\beta}
\newcommand{\gc}{\gamma}
\newcommand{\gd}{\delta}
\newcommand{\gep}{\epsilon}
\newcommand{\gee}{\varepsilon}
\newcommand{\gz}{\zeta}
\newcommand{\get}{\eta}
\newcommand{\gth}{\theta}
\newcommand{\gthh}{\vartheta}
\newcommand{\gi}{\iota}
\newcommand{\gk}{\kappa}
\newcommand{\gl}{\lambda}
\newcommand{\gm}{\mu}
\newcommand{\gn}{\nu}
\newcommand{\gks}{\xi}
\newcommand{\go}{\0}
\newcommand{\gp}{\pi}
\newcommand{\gpp}{\varpi}
\newcommand{\gr}{\rho}
\newcommand{\grr}{\varrho}
\newcommand{\gs}{\sigma}
\newcommand{\gss}{\varsigma}
\newcommand{\gt}{\tau}
\newcommand{\gu}{\upsilon}
\newcommand{\gf}{\varphi}
\newcommand{\gff}{\varphi}
\newcommand{\gx}{\chi}
\newcommand{\gps}{\psi}
\newcommand{\gw}{\omega}
\newcommand{\gG}{\Gamma}
\newcommand{\gD}{\Delta}
\newcommand{\gTh}{\Theta}
\newcommand{\gL}{\Lambda}
\newcommand{\gKs}{\Xi}
\newcommand{\gP}{\Pi}
\newcommand{\gS}{\Sigma}
\newcommand{\gU}{\Upsilon}
\newcommand{\gF}{\phi}
\newcommand{\gPs}{\Psi}
\newcommand{\gW}{\Omega}

\newcommand{\ti}{\tilde}
\newcommand{\Li}{{\cal L}}
\newcommand{\ra}{\rightarrow}
\newcommand{\pa}{\partial}
\newcommand{\ov}{\overline}
\newcommand{\fad}{\frac{\Delta T}{T}}
\newcommand{\lan}{\langle}
\newcommand{\ran}{\rangle}

\begin{abstract}
 We consider the simple inflationary model with  peculiarity in the form of 
"plateau" in the inflaton potential. We use the formalism of coarse-grained field in order
to describe the production of metric perturbations $h$ of an arbitrary amplitude,
and obtain non-Gaussian probability function for such metric perturbations. 
We associate the  spatial regions having large perturbations $h\sim 1$
with the regions going to primordial black holes after inflation. 
We show that 
in our model the non-linear effects can lead to overproduction of the primordial 
black holes.

PACS number(s): 98.80.Cq, 97.60.Lf, 98.70. Vc, 98.80.Hw
\end{abstract}

\section{Introduction}
Starting from pioneering works by Zel'dovitch and Novikov [1], and also
by Hawking [2], the primordial black holes (hereafter PBH's) were subject
of extensive ivestigations. The presence of PBH's 
may significantly influence on physical processes and effects 
in the Universe (such as nucleosynthesis, CMBR spectral distortions, 
or distortions of $\gamma$-ray background radiation)
due to Hawking effect [3], PBH's
may be a component of dark matter (see e.g. [4], [5]).
The formation of PBH's is determined by small scale, but large amplitude
inhomogeneities in the Early Universe, and the processes 
of PBH's formation, evolution and decay link the physical conditions of
Early Universe with conditions in the radiation-dominated epoch
and present-day cosmology. Even the very absence of PBH's may significantly 
constraint the models of the beginning of cosmological evolution.

Usually the processes of PBH's formation are associated with production of the scalar mode
of perturbations during inflation
(see e.g.  [5-9]) or phase transitions in the Early Universe [10]. 
In this paper
we are going to discuss the first possibility, which allows to use
the powerful and well-elaborated theory of instability
of the expanding
Universe for analysis of conditions, under which PBH's can form.

The theory of generation of adiabatic perturbations during inflation started 
from pioneering papers [11-13].
It was established that the RMS-amplitude of metric perturbations 
$\delta_{rms}$ is connected with the parameters of inflationary theory
by means of relation
$$\delta_{rms}={1\over 2\pi}{H^{2}\over |\dot \phi|}, \eqno 1$$
where $H$ is the Hubble parameter, $\dot \phi$ is the velocity of the
field, evolving during inflation.
To get PBH's abundance
in an observable amount, one should have $\delta_{rms}\sim 10^{-2}-10^{-1}$
(see, e. g. [14]).On the other hand COBE CMBR data, as well as analysis of
Large-Scale Structure formation constraint the amplitude of perturbations 
$\delta_{rms}\sim 10^{-5}$ at super-large scales. Therefore to get PBH's
one should increase the amplitude of the perturbations by a factor 
$10^{3}-10^{4}$ at small scales. Unfortunately this cannot be reached in 
the simplest inflationary models, since in these models $\delta_{rms}$
logarithmically grows with increase of scale, and one should use
nonstandard models having additional power at small scales
to obtain significant PBH's amount.

Recently, several models of such type were proposed. For instance,
Carr and Lidsey [6] proposed toy model having blue type spectrum
(the spectrum $\delta_{rms}(k)\propto k^{a}$, where $k$ is 
the wavenumber, and $a$ is the spectral index), and investigated
the constraint on the spectral index $a$ associated with possible 
PBH's formation in such model.
Linde [15] has shown that blue type spectra can be naturally obtained
in the two-field model of so-called hybrid inflation.

Another type of model having a spike in the power spectrum at some scale $k_{bh}$
was proposed by Ivanov, Naselsky and Novikov ([5], hereafter INN)
\footnote{See also the papers by Hodges and Blumenthal, Hodges 
{\it et
all}[16] and Kates 
{\it et
all} [17], who employed similar models in contest of Large-Scale Structure
formation theory} . 
They considered one-field inflationary model with 
inflaton $\phi$ and assumed that 
the potential $V(\phi)$ has 
a "plateau" region at some scale $k_{bh}$, and has a
standard form (say, power-law form) outside the "plateau" region.
The field $\phi$ slows down in the "plateau" region giving increase of 
the spectrum of perturbations at the scale $k_{bh}$
 according to eqn. $(1)$. One can
adjust the parameters of "plateau" region to obtain the desired increase of
the spectrum,
and consequently the desired PBH's amount. Garcia-Bellido {\it et all} [8]
and also Randall {\it et all} [9] considered more realistic two field models
having a saddle point in two-dimensional form of potential $V(\phi, \psi)$.
Like the one-field model, the evolution of the system 
of fields slows down near the saddle point giving an increase of the spectrum
power. Randall {et all} pointed out that such models solve several
fine-tuning problems of the standard inflation, and therefore look
very natural from the point of view of high energy physics. Garcia-Bellido {\it et
all} carefully investigated the process of PBH's formation in such models
(see also recent work by Yokoyama, [18]).  

If the primordial black holes are not super-large,
they probably collapse during the radiation dominated epoch of the evolution 
of the Universe. This means that the amplitude $h_{*}$ of the metric inhomogeneities 
inside the regions going to PBH's should be of order of unity to overcome the
strong pressure forces during collapse of the perturbed region [14].
These large amplitude metric inhomogeneities are assumed to be generated
during inflation as rare events in the random field of the metric perturbations.
Since the amplitude of the inhomogeneities $h_{*}$ is rather large,
the natural question appears: to what extent we can rely on the linear theory
of perturbations which usually gives Gaussian probability distribution
of PBH's formation?     

To answer this question we can apply the formalism of coarse-grained fields
(introduced by Starobinsky [19]) as an alternative approach to the linear
theory that can describe large amplitude deviations 
of the field and the metric from the background quantities. 
According to this approach, the spatially inhomogeneous field 
$\phi(\vec x, t)$ is divided into two parts: the large-scale part 
$\phi_{ls}$, which consists of the modes with physical wavelengths
$\lambda \propto ak^{-1}$ greater than some characteristic scale
$\lambda_{c-g} \ge H^{-1}$, and the small scale part which consists of modes
with $\lambda < \lambda_{c-g}$. During inflation, the physical wavelengths
are stretched, and new perturbations are added to $\phi_{ls}$.  
This effect may be considered as a new random force $f(t)$ in the equation of motion of 
the field $\phi_{ls}$, and usually the dynamics of $\phi_{ls}$ is described
in terms of diffusion equation for probability density $\Psi(\phi_{ls}, t)$.
This equation was subject of a number of works in connection with problems
of Quantum Gravity and Large-Scale Structure formation. Recently, it was pointed
out, that this equation can be employed for calculations of the probability to
find large amplitude peaks in the random distribution of field $\phi_{ls}$,
and it was mentioned that such approach can be applied to the problem of
PBH's formation [20].

Here we would like to note that when studying the effects originating after the end 
of inflation, such as PBH's formation, one should use the large scale part of metric
instead of large scale part of field.
Contrary to
the field $\phi_{ls}$, the large scale part of the metric, namely the "inhomogeneous  
scale factor $a_{ls}(\vec x)$" (see eqns. $(23-24)$ for exact definition)
is the quantity conserving during the evolution outside the horizon,
and this property allows to
connect the physical conditions during inflation with the physical conditions
during radiation-dominated epoch, when PBH's are formed. Moreover,
the criterion for PBH's formation can be directly formulated in terms of
$a_{ls}(\vec x)$ ( Refs. [21], [22]). 
Therefore, the calculation of $a_{ls}(\vec x)$ 
gives a tool to describe quantitatively the generation of non-linear metric
perturbations, and the evolution of these perturbations into PBH's.

In this paper we calculate the probability distribution 
function ${\cal P}(a_{ls}(\vec x))$ in the model with almost flat region 
in the inflaton potential. The main idea of our calculations has already been applied
in the models of so-called stochastic inflation (see, e. g. [23] and references
therein), and is very simple. When the field $\phi_{ls}$ evolves inside the plateau
region it slows down, and the random kicks (described by the force $f(t)$) 
significantly  influence on its evolution. So, the trajectory of the field inside 
the plateau region becomes stochastic, and the time $\Delta t$ that the field spends
on the plateau, depends on the realization of the stochastic process. 
The total increase
of the scale factor $a_{ls}$ during the field evolution on the plateau, is obviously
determined by $\Delta t$: $a_{ls}\propto e^{H\Delta t}$.  Since different regions
of the Universe separated by distances greater than $H^{-1}$ evolve independently,
the increase of $a_{ls}$ corresponding to different regions is determined by 
different realizations of
the random process. Thus the scale factor $a_{ls}$ varies from one region to another     
after the field passes the plateau, that is the quantum effects generate the
coordinate dependence of the scale factor. The shape of function $a_{ls}(\vec x)$ is
conserved during the subsequent evolution of the Universe until the scale of inhomogeneity 
crosses horizon at the second time. At that time, in the regions with significant contrast
of $a_{ls}(\vec x)$ the primordial black holes are formed.

Using the approach described above we calculate the probability distribution function
${\cal P}(a_{ls}(\vec x))$. With help of a simple criterion of PBH's formation we
relate ${\cal P}(a_{ls}(\vec x))$ to the probability  of PBH's formation. We show that
in our case the non-linear effects over-produce PBH's
\footnote{Note that this result differs from that obtained in Ref. [20].}.
Although this result is very important qualitatively, it does not significantly
change the estimate based on the linear theory.

We use the simple one-field model, proposed by INN (see also Refs. [24], [25]). Due to
simplicity of this model the  bulk of our results are obtained analytically.
We hope that our approach provides a reasonable approximation to the case of 
more complicated two-field models. We are going to discuss these models in our future
work.
   
The paper is organized as following. We introduce our model 
and discuss the classical dynamics of the metric and field in Section 2. 
In Section 3 we obtain an expression for
${\cal P}(a_{ls}(\vec x))$. We consider the role of non-linear effects on the
statistics of PBH's production in Section 4. We summarize our conclusions and 
discuss applicability of our approach in Section 5.

\section{The dynamics of classical model}

In this Section we consider the classical dynamics of spatially
homogeneous 
parts of metric and field in the simplest inflationary model with
a single scalar field (inflaton) and with the peculiarity in the inflaton potential. 
In this case the system of
dynamical equations contains only two dynamical variables
-scale factor $a(t)$ and spatially homogeneous part $\phi_{0}(t)$ of the
field
$\phi$, and reduces to the Hamiltonian constraint
equation
$$H^{2}={8\pi\over 3}(V(\phi_{0})+{{\dot \phi_{0}}^{2}\over 2}), \eqno 2$$
and to the equation of motion for field $\phi_{0}$
$$\ddot \phi_{0} +3H\dot \phi_{0} +{\partial \over \partial \phi}V(\phi_{0})=0,
\eqno 3$$  
where $H={\dot a\over a}$, and another symbols have their usual meaning.
We hereafter use the natural system of units.

We assume that 
the effective potential $V(\phi)$ has a small
almost flat region ('plateau') between some characteristic values of 
field $\phi_{1}$ and $\phi_{2}$. The
potential is also assumed to be proportional to $\phi^{4}$ 
outside the 'plateau' region
$$V(\phi)={\lambda \phi^{4}\over 4} \eqno 4$$
at $\phi < \phi_{1}$
$$V(\phi)=V(\phi_{1})+A(\phi -\phi_{1}) \eqno 5$$
at $\phi_{1}<\phi <\phi_{2}$, and
$$V(\phi)={\tilde \lambda \phi^{4}\over 4} \eqno 6$$
at $\phi > \phi_{2}$. Here $V(\phi_{1})={\lambda \phi_{1}^{4}\over 4}$,
$\tilde \lambda =\lambda {({\phi_{1}\over \phi_{2}})}^{4}+{4A(\phi_{2}
-\phi_{1})\over \phi_{1}^{4}}$. As we will see
below the size of the flat region 
is very small $\Delta \phi=\phi_{2}-\phi_{1} \ll \phi$,
${A(\phi_{2}-\phi_{1})\over V(\phi_{1}}\ll 1$ so we can set
$\lambda \approx \tilde \lambda$.   
 At sufficiently large values of $\phi_{0} > 1$
the kinetic term in the equation $(2)$ is negligible in comparison with
the potential term 
$${{\dot \phi_{0}}^{2}\over 2} \ll V(\phi_{0}), \eqno 7$$
and the equation $(2)$ reduces to an algebraical
relation between $H$ and $\phi_{0}$ (so-called slow-roll approximation)
$$H=\sqrt{{8\pi \over 3}V(\phi_{0})}. \eqno 8$$
From the equation $(8)$ it follows that the Universe expands
quasi-exponentially ($H\approx const$ and $a\propto e^{Ht}$) at 
$\phi_{0} > 1$.

It can also be easily shown that outside the plateau region the field moves
with large friction at $\phi_{0} > 1$, so
$$|\ddot \phi_{0}|\ll |3H\dot \phi_{0}|. \eqno 9$$
The friction dominated condition $(9)$ helps to simplify the integration
of the system $(2-3)$. Integrating the eqns. $(2-3)$ with help of inequalities
$(7)$, $(9)$ at $\phi_{0} > \phi_{2}$, we have
$$\phi_{0}(t)=
\tilde \phi_{0}\exp-{(\sqrt{{\tilde \lambda \over 6\pi}}t)}, \eqno 10$$   
and
$$a(\tilde \phi_{0})=a_{0}\exp{(N(\tilde \phi_{0})-N(\phi_{0}))}, \eqno 11$$
where $\tilde \phi_{0}$ and $a_{0}$ are some initial values of the field 
and scale factor. 
$$N(\phi_{0})=\int_{\phi_{2}}^{\phi_{0}} Hdt=
\pi(\phi_{0}^{2}-\phi^{2}_{2}) \eqno 12$$
is the number of e-folds of the scale factor during the 
field rolling down starting from some initial
value of $\phi$ and down to the field $\phi_{2}$.
The similar formulae hold at $\phi_{end} < \phi_{0} < \phi_{1}$
$$\phi_{0}(t)=
\phi_{1} \exp(-\sqrt {{\lambda \over 6\pi}}(t-t_{1})), \eqno 13$$
$$a(\phi_{0})=a_{1}\exp (N_{end}(\phi_{1})-N_{end}(\phi)), \eqno 14$$
where $\phi_{0}(t_{1})=\phi_{1}$ and $a_{1}=a(t_{1})$, and
$N_{end}(\phi_{0})$ is the number of e-folds up to the end of
inflation: $N_{end}(\phi_{0})=\pi (\phi_{0}^{2}-\phi_{end}^{2})$, 
where we assume that inflation ends at standard (for $\lambda \phi^{4}$
theory)
value of $\phi_{end}={1\over
\sqrt{2\pi}}$. Note, that $N_{end}(\phi_{1})$ should be rather large.
For example, to get a feature in the spectrum at scales, corresponding to the
solar mass, we should have $N_{end}(\phi_{1})\sim 50-60$. Therefore,
 the value of
$\phi_{1}$ should be greater than unity ($\phi_{1}\sim 4.5$ for 
$N_{end}(\phi_{1})\sim 60$). 

Now let us consider the dynamics of inflaton in the "plateau" region $\phi_{1}
< \phi_{0} < \phi_{2}$. In this region the equation $(3)$ is simplified to
$$\ddot \phi_{0} +3H_{0}\dot \phi_{0}+A=0, \eqno 15$$
where $H_{0}=\sqrt{{8\pi\over 3}V_{0}}$. The solution of eqn. $(15)$ can be
written as
$$\phi_{0} = \phi_{2} +{1\over 3H_{0}}\dot \phi_{in}(1-e^{-3H_{0}t})
-{At\over 3H_{0}}
=\phi_{2}-{1\over 6\pi \phi_{2}}(1-e^{-3H_{0}t})-{At\over 3H_{0}}, \eqno 16$$
and for the field velocity we have
$$\dot \phi_{0}=\dot \phi_{in}e^{-3H_{0}t}-{A\over 3H_{0}}, \eqno 17$$
where 
$\dot \phi_{in}=\dot \phi_{0}|_{\phi_{0}=\phi_{2}}
=-{1\over 3H_{0}}{\partial \over
\partial \phi}V(\phi_{2})=-\sqrt{\tilde \lambda \over 6\pi}\phi_{2}$
is the field velocity at the moment $t$=0
of entrance of the field in the "plateau"
region. The second term in the eqn. $(16)$ and the 
first term in the eqn. $(17)$ are due to inertial influence of initial velocity
$\dot \phi_{in}$, and the last terms in the both equations are due to nonzero
slope of potential in the plateau region. The evolution of the field in the
plateau region can be divided into two stages. At first stage the field evolves
mainly due to inertial term, and velocity exponentially decreases with time.
After some characteristic time $t_{*}$ the nonzero slope of potential
$A$ starts to determine the evolution, the velocity tends to the constant value
$\dot \phi_{fd}=-{A\over 3H_{0}}$, and the field amplitude starts to decrease
linearly with time. The time $t_{*}$ can be estimated by equating the inertial
and potential terms in the eqn. $(16)$, and is determined by the condition
$3H_{0}t_{*}e^{3H_{0}t_{*}}={B\over A}$, where 
$B={\partial \over \partial \phi}V(\phi_{0}=\phi_{1})$.
As we discussed in Introduction, the spectrum amplitude is inversely proportional
to the field velocity 
$(\delta_{rms}\approx {1\over 2\pi}{H^{2}\over |\dot \phi|})$, therefore we need
to slow down the velocity approximately by $\sim 10^{3}-10^{4}$ times 
to get the 
increase of the spectrum amplitude from the initial value $\delta_{rms}(in)=
{1\over 2\pi}{H^{3}\over B}\sim 10^{-5}$ up to the typical for PBH production
$\delta_{rms}\sim 10^{-2}-10^{-1}$. 
For that, we should fix the "amplification"
parameter $\alpha ={B\over A}\sim 10^{3}-10^{4}$. 

Our model has two possible limiting variants depending on the relation between
the time $t_{c}$ of the crossing of plateau region by the field $\phi_{0}$ 
($\phi_{0}(t_{c})=\phi_{1}$) and 
$t_{*}$. If $t_{c}\approx t_{*}$ the field crosses the plateau mainly due to
inertia. In this case the parameter $\alpha$ determines the number of e-folds
during plateau crossing $\delta N \approx H_{0}t_{c}\approx {1\over 3}\ln \alpha
\approx 2.3$, and therefore the width of produced bump in the spectrum 
remains small and fixed. The model of similar type was discussed by INN. 
Here we consider another possible case
$t_{c} > t_{*}$, where the field spends some time on the plateau, evolving in the
friction-dominated approximation. In this case the width of the spectrum
is determined by the value of $t_{c}$, which is the free parameter of our model.     
Instead of $t_{c}$ we will parameterize our model by the quantity $\gamma$-the
ratio of wave numbers, corresponding to the fields $\phi_{1}$, $\phi_{2}$, 
respectively, $t_{c}=H_{0}^{-1}\ln \gamma$. 
The parameter $\gamma$ cannot be too 
small $\gamma > \alpha^{1/3}$ and we take $\gamma \approx 10^{3}$ in the
estimations. 
If $\gamma$ is not extremely large $\ln \gamma \ll N(\phi_{1})$, 
the size of plateau $\Delta \phi_{0}=\phi_{2}-\phi_{1}$ is of 
order of typical size $\Delta \phi_{*}={B\over 9H^{2}}$. The typical
relative size of plateau is very small   
$${\Delta \phi_{0} \over \phi_{0}} =
{1\over 6\pi \phi_{0}^{2}}\approx{1\over 6N(\phi_{1})}
\approx 0.003. \eqno 18$$
Thus, the correction due to the presence of plateau 
practically does not influence
on the dynamics of the field outside plateau region and we can set $\lambda
=\tilde \lambda$. On the other hand, the size of plateau is much greater than
$H_{0}$ - the typical size of quantum fluctuations,
$\Delta \phi_{*}={H_{0}\over 6\pi \delta_{rms}(in)} \sim 10^{5} H_{0}$.   
 
Typically, the estimate ${\Delta \phi_{0} \over \phi_{0}} \ll 1$ 
holds for arbitrary
power-low potentials $V(\phi) \propto \phi^{p}$ provided power $p$ is not
very large. However the opposite limiting case is also possible. For example,
Bullock and Primack [20] proposed the potential of the form
$$V(\phi)=\lambda_{bp}(1+\arctan{(\phi)}),\quad \phi > 0$$
$$V(\phi)=\lambda_{bp}(1+4*10^{33}\phi^{21}),\quad \phi < 0 \eqno 19$$ 
where the  constant $\lambda_{bp}=6*10^{-10}$ is chosen to normalize the 
large-scale part of spectrum to the RMS-amplitude $\approx 3*10^{-5}$.
The flat region in this potential starts from $\phi =0$ and ends at $\phi=
-1.23*10^{-2}$, and inflation ends itself at $\phi=\phi_{end}=-1.55*10^{-2}$.
It was mentioned by Bullock $\&$ Primack that this potential leads to strongly
non-Gaussian statistics of field perturbations.

\section{Non-linear metric perturbations 
from the quantum dynamics of coarse-grained field}

It is well known that there are two equivalent ways to describe inhomogeneous
Universe. The first way is to consider inhomogeneities as a small corrections
to the homogeneous space-time and study them in the frameworks of linear
theory of perturbations. Another approach splits the metric and the field
into large-scale part (coarse-grained over some scale greater than
horizon scale), and small-scale part. During inflation, the dynamical 
equations for coarse-grained field $\phi_{ls}$ and 
coarse-grained scale-factor $a_{ls}$ are equivalent to eqns. $(3,8)$ provided
the quantum effects are switched off. The quantum effects continuously produce
new inhomogeneities of random amplitude with scales greater than the scale of
coarse-graining. These inhomogeneities should be added to $\phi_{ls}$ and $a_{ls}$
and effectively this leads to the presence of stochastic force term in the equations 
of motion. Therefore, the dynamics of coarse-grained variables can be described
in terms of the distribution functions of $\phi_{ls}$ and $a_{ls}$, and in principal
these distribution functions can provide the same information as the power
spectrum of perturbations, and furthermore the coarse-grained formalism gives a
tool for description of the metric perturbations with amplitude, greater than 1. 

The effective dynamical equation for the field $\phi_{ls}$ has the form
[19]
\footnote{See also recent papers [26] and references therein.}
$$\ddot \phi_{ls} +3H_{ls}\dot \phi_{ls} +{\partial \over \partial \phi}V(\phi_{ls})
=D^{1/2}f(t), \eqno 20$$
where $D={9H_{ls}^{5}\over {(2\pi)}^{2}}$, and 
$f(t)$ is delta-correlated random 
force, $<f(t_{1})f(t_{2})>=\delta (t_{1}-t_{2})$. 
The equation for coarse-grained scale factor $a_{ls}$ remains 
unchanged
$$H_{ls}=\sqrt{ {8\pi \over 3}V(\phi_{ls})}. \eqno 21$$
The solution of the set of eqns. $(20,21)$ is extremely difficult problem, and
can be done under some additional simplifying assumptions. For example if we 
choose the featureless potential, and consider
the friction-dominated solutions of the eqn. $(20)$, we can obtain the 
solutions  describing self-reproduced inflationary Universe (provided
the stochastic term in $(20)$ dominates over potential term, see for
example Linde [23]).
In our
case we cannot use the friction-dominated condition in the beginning
of the field evolution inside the plateau
region. However we can adopt another simplifying assumptions: first 
we can set $H_{ls}=H_{0}=const$ inside and near the plateau region, 
and second, we can omit 
the stochastic term in the eqn. $(20)$ outside the plateau region, assuming the
field moves along the classical trajectory there. 
Under these assumptions the statistics of the scale factor $a_{ls}$ is
totally determined by the time $\Delta t$ that field $\phi_{ls}$ spends
in the plateau region
$$\Delta N=ln(a_{out}/a_{in})=H_{0}\Delta t, \eqno 22$$
where $a_{in}$ is the value of scale factor at the time $t=0$
of entrance of the field in the plateau region,
and $a_{out}$ corresponds to the moment $\Delta t$, when the field leaves
the plateau region. 
To see that let us consider the evolution of the scale factor $a_{ls}$ in the
comoving coordinate system. 
Outside the horizon the hypersurfaces of constant comoving time $t_{com}$ 
practically coincide with hypersurfaces of constant energy density
$\epsilon =const$. On the other hand, the field $\phi_{ls}$ evolves
slowly during inflation and hypersurfaces of constant energy density
are close to hypersurfaces $\phi_{ls}=const$, and therefore we can
put $a_{ls}(t_{com})=a_{ls}(\phi_{ls})$.
After the field passes the plateau region,
the evolution of $a_{ls}(\phi_{ls})$ can be described by the standard 
expression $(14)$, so we have 
$$a_{ls}(\phi_{ls})
=a_{in}\exp{(\pi(\phi_{1}^{2}-\phi_{ls}^{2})+\Delta N)}, \eqno 23$$
where $\Delta N$ is nearly constant inside of coarse-grained regions
with comoving scale $\lambda_{c-g} \approx a_{out}H_{0}^{-1}$, but changes 
from one
region to another. Thus, the metric outside horizon has 
the quasi-isotropic form
$$ds^{2}=dt^{2}-a^{2}_{ls}(\phi_{0})
a_{ls}(\vec x)\delta^{i}_{j}dx_{i}dx^{j}, \eqno 24$$
where we represent the scale factor $a_{ls}(\phi_{ls})$ as a multiplication
of two factors: $a(\phi_{0})$ and $a_{ls}(\vec x)\equiv e^{\Delta N}$. 
Here $a_{ls}(\phi_{0})$ and $\phi_{0}(t)$ are 
determined by the classical equations $(13)$, $(14)$, 
and the spatial coordinates $\vec x$ are coarse-grained over the regions
with scale $\lambda_{c-g}$.
To estimate the change of metric from one region to another quantitatively, 
we introduce 
the definition of non-linear metric perturbation
$$h\equiv {a_{ls}(\phi_{ls}) -a(\phi_{0})\over a(\phi_{0})}=
\exp H_{0}{(\Delta t -t_{c})} -1 \eqno 25$$  
(remind, that $t_{c}=H_{0}^{-1}\ln{\gamma}$ is the time which the field
spends in the plateau region moving along the classical trajectory when
the stochastic term in $(20)$ is switched off). Note, that in the 
limit of small $h \ll 1$, the metric assumes the form
$$ds^{2}=dt^{2}-a^{2}(\phi_{0})(1+2h(\vec x))\delta^{i}_{j}dx_{i}dx^{j}, 
\eqno  26$$ 
and the definition $(25)$ is reduced to the standard expression for 
growing mode of adiabatic perturbation outside the horizon.
Namely, in this case $h$ reduces to gauge independent
quantities, introduced by a number of authors
[11-13], [27] up to a constant factor.
The variables $(25, 26)$ do not depend on time outside the horizon.
Therefore, using of these variables is very convenient to match
the perturbations, generated during inflation with the perturbations,
crossing horizon at the normal stage
of the Universe evolution. 
As one can see from $(25)$ the metric perturbations
are determined by stochastic variable $\Delta t$ and the distribution of 
$\Delta t$ must follow from the solution of eqn. $(20)$. Note, that 
the definition of non-linear metric perturbations should be taken
with a caution. In principal, one can use another definition relating
to $(25)$ by some non-linear transformation, and having the same     
limit $(26)$ in the case of small $h$. For example,
Bond and Salopek [28] used the quantity 
$\tilde h=\ln{({a_{ls}(\phi_{ls})\over a(\phi_{0})})}$ to define
non-linear metric perturbations. However, the criterion
for PBH's formation can be directly expressed in terms of 
the quantity $(25)$ (see next Section), and therefore this quantity
is the most natural variable for our purposes.

Although the assumption of constant $H_{0}$ greatly simplifies the
problem it still remains rather complicated for a simple
analytical treatment
\footnote{In this case our problem is reduced to the first-passage problem 
for one-dimensional Fokker-Plank (Kramers) equation, associated with
eqn. $(20)$ [29]. The general solution of this problem demands too much formalism
[30], and is not considered here.  Note, however that the simple asymptotic estimates are 
still possible in this case  [30], [31].}

For further progress we have to make some additional
assumptions. We will consider below the plateau region of sufficiently
large size. 
For this case the field approaches to the end of plateau
in the friction-dominated approximation, which greatly simplifies the
treatment of diffusion processes. To estimate the relevance of 
friction-dominated approximation we should compare the time 
$t_{c}$ and the time $t_{*}\sim \ln{(\alpha)}$ of 
the decay of the inertial
term $\ddot \phi$ in the eqns. $(15-17,20)$.
 If $t_{c} > t_{*}$ and therefore $\gamma \gg \alpha^{1/3}$, 
the inertial term in these equations can be neglected
at $t_{*} < t < t_{c}$. In this regime the solution of 
the classical equation of motion $(15)$ has the form
$$\phi_{0}(\tau)\approx \phi_{2} -a\tau, \eqno 27$$
and the equation $(20)$ becomes
$${d\phi_{ls} \over d\tau} + a=d^{1/2}f(\tau), \eqno 28$$
where $\beta=3H_{0}$, and we introduce the dimensionless 
time $\tau=\beta t$, $a=A/\beta^2$
and $d={D\over 2\beta^{3}}={H^{2}_{0}\over
24\pi^{2}}$. The stochastic equation $(28)$ is associated with 
simple diffusion type equation, describing the evolution of positions
probability distribution $\Psi (\tau, \phi)$
$${\partial \Psi \over \partial \tau}=
d{\partial^{2} \over \partial \phi^{2}}\Psi 
+a{\partial \over \partial \phi}\Psi, \eqno  29$$
Now we assume that the distribution $\Psi$ is not
spread out sufficiently
before $\tau_{*}=\beta t_{*}$ and take $\delta$-distributed $\Psi$ function
at the moment $\tau =\tau_{*}$ as the initial condition for our problem
$$\Psi(\tau_{*})=\delta (\phi_{ls} -\phi_{*}), \eqno 30$$
where $\phi_{*}=\Delta \phi -a\tau_{*}$ is the value of field corresponding
to the beginning of "friction-dominated" part of plateau region
\footnote{The estimates show that the characteristic width of 
$\Psi(\tau_{*})$ is of order of $H$ and much less than the size
of "friction-dominated" region $\phi_{*}-\phi_{1}$.}.
 
Together with initial condition $(30)$ we should specify the boundary
condition at $\phi_{ls}=\phi_{1}$. This condition depends on the form of 
the transition layer between the plateau region and the part of potential with
steep slope ${\partial \over \partial \phi}V(\phi)=B$. 
We assume this transition to be sharp, and therefore set the condition of
absorbing wall at the downstream point $\phi_{ls}=\phi_{1}$
$$\Psi (\phi_{1},\tau)=0, \eqno 31$$
Note, that this boundary condition was used by Aryal and Vilenkin [32]
for analysis of stochastic inflation in the theory with top-hat
potentials. In that paper it was shown that the more reasonable
smooth transitions between the flat and steep regions of the potential
are unlikely to modify significantly the resulting distribution.

In our case the probability density ${\it P}(\tau)$ of time $\tau$ 
relates to the solution of eqn. $(27)$ as
$${\cal P}(\tau)=S|_{\phi_{ls}=\phi_{1}}=d{\partial \over \partial \phi}\Psi,
\eqno 32$$
where we define by $S$ the probability 
current $S=d{\partial \over \partial \phi}\Psi +
a\Psi$. 
The conservation of the probability current allows 
to estimate the correction term to eqn. $(32)$ due to 
nonzero $\Psi (\phi_{1})$.
Assuming that field moves along the classical trajectory after 
$\phi_{ls}=\phi_{1}$,
we have 
$S(-\phi_{1})\approx {B/\beta^{2}}\Psi \approx S(+\phi_{1})
\approx d{\partial \over \partial \phi}\Psi$. Therefore the correction to the
expression $(32)$ is ${\beta^{2} a\over B}=\alpha^{-1}\sim 10^{-3}-10^{-4}$ 
times smaller than the leading term. 

The conditions $(30,31)$ determine the solution of eqn. $(29)$. 
This solution   
can be found by standard methods of the theory of 
diffusion equations (see, e. g. Ref. 
[33]), and in our case has the form 
$$\Psi(\phi, \tau)={1\over \sqrt{4\pi d(\tau -\tau_{*})}}\exp{\lbrace 
-{1\over 4d(\tau-\tau_{*})}
{(\phi-\phi_{*}+a(\tau-\tau_{*})}^{2}\rbrace} $$
$$(1-\exp\lbrace -{1\over d(\tau
-\tau_{*})}(\phi-\phi_{1})(\phi_{*}-\phi_{1})\rbrace ), \eqno 33$$
Substituting $(33)$ to the equation $(32)$ we find the explicit
expression for ${\cal P}(\tau)$
$${\cal P}(\tau)={1\over \sqrt{(4\pi d(\tau -\tau_{*})}}({\phi_{*}-\phi_{1}\over
\tau -\tau_{*}})\exp {\lbrace -{1\over 4d(\tau -\tau_{*})}{(\phi_{1}-\phi_{*}
+a(\tau-\tau_{*}))}^{2}\rbrace }. \eqno 34$$ 
The expression for probability distribution of metric can be readily obtained from
$(34)$. Using the eqn. $(22-25)$ to express the time $\tau$ in terms of $h$,
taking into account the eqn $(27)$ and the definitions of $a$, $d$, 
and assuming $a>0$, we obtain
$${\cal P}(h)={1\over \sqrt {2\pi \delta^{2}_{pl}}}{N_{cl}
\over {N_{st}}^{3/2}}{dN_{st}\over dh}\exp {\lbrace -{{(N_{st}-N_{cl})}^{2}\over
2\delta^{2}_{pl}N_{st}}\rbrace }, \eqno 35$$
where $\delta_{pl}={3H_{0}^{3}\over 2\pi A}=\alpha\delta_{rms}(in)$ is the standard metric
amplitude calculated for the plateau parameters, and 
$$N_{cl}=\ln{\gamma}-\tau_{*}/3, \quad 
N_{st}=\ln{(1+h)} +N_{cl}\eqno 36$$
are the numbers of e-folds 
for the classical path $\phi_{0}(t)$ and for
a random path $\phi_{ls}(t)$, which start
at $\phi_{*}=\phi(t_{*})$ and end at $\phi_{1}$. 
 
When the perturbations are small $N_{st}-N_{cl}\approx h \ll 1$, 
the distribution $(35)$ has the standard Gaussian form
$${\cal P}(h)={\it P}_{G}(h)={1\over \sqrt{2\pi \delta_{pl}^{2}N_{cl}}}
\exp{ \lbrace -{h^{2}\over 2\delta_{pl}^{2}N_{cl}}\rbrace }, \eqno 37$$
and in the opposite case of very large metric perturbations
$h \gg 1$ and $N_{st}\sim \ln h > N_{cl}$ 
the distribution ${\cal P}(h)$ deviates sharply from the 
Gaussian law and has the power-law form  
$${\cal P}(h) \propto h^{3/2+\delta_{pl}^{-2}/4}, \eqno 38$$

As seen from eqns. $(35-38)$,  
the non-Gaussian effects over-produce the 
metric perturbations of high amplitude in our model. 
To understand this fact, let us discuss the origin of non-Gaussian effects in our
model. There are two sources for such effects. First, note that the "effective
dispersion" $\sigma^{2}_{eff}=\delta_{pl}^{2}N_{st}$ in eqn. $(35)$ depends itself
on the value of the stochastic variable $N_{st}$. Qualitatively, 
it can be explained as 
follows. In the linear theory the dispersion $\sigma^{2}=\delta^{2}_{pl}N_{cl}$ is
proportional to the time spent by the classical background field $\phi_{0}$ on the 
plateau. In non-linear theory the coarse-grained field $\phi_{ls}(t)$ plays the role
of background field, 
and therefore the distribution of the family of neighboring to $\phi=\phi_{ls}(t)$
paths should be described in terms of the probability distribution with dispersion
$\sigma^{2}_{eff}$, which is proportional to the time spent by 
field $\phi_{ls}$ on the plateau.  Second,
the amplitude of large metric perturbations $h$ depends on $N_{st}$
exponentially ($h\sim e^{N_{ls}}$), so order of magnitude increase of $N_{st}$
leads to exponential increase of $h$. Obviously, these two effects increase the
probability of large amplitude metric perturbations. 

\section{Probability of black holes formation}

Although the distribution $(35)$ provides very important information
about the geometry of spatial part of metric outside horizon, 
it cannot be directly applied
to the estimates of PBH's formation. Indeed, the distribution $(35)$
is formed by the field inhomogeneities with wave-numbers $k$ in the
range $(\Delta k=[k_{min}\approx a_{in}H_{0} < k < k_{max}\approx a_{out}H_{0}])$. 
The process of PBH formation is determined mainly by
the field modes with wave-numbers $(\delta k \approx k_{bh} \ll \Delta k)$,
where $k_{bh}$ is the typical PBH wavenumber. The modes with $k < k_{bh}$
compose the large-scale background part of metric
at the moment of PBH formation, and 
do not influence on the formation of PBH's significantly. 
The modes with $k > k_{bh}$ lead to
high-frequency modulation of the perturbation with $k\sim k_{bh}$,
which is also unimportant, provided the mode with $k\sim k_{bh}$
crosses the horizon second time 
at the radiation-dominated epoch. Therefore, 
in order to obtain the probability of
PBH's formation, we should subtract the contribution
of the large-scale and small-scale metric perturbations.

In general it is very difficult to separate the perturbations of a given
scale in the frameworks of non-linear approach. However, we can estimate
the probability density of the perturbations, corresponding
to the smallest scale $k_{bh}\approx a_{out}H_{0}$  
\footnote{In this connection, let us note that the black holes of smallest
mass should give the major contribution to the present fraction of
black holes, provided PBH's spectrum is flat (Carr, 1975 [14]).}. 
For that we simply
put $N_{cl}=1$ in eqns. $(35,36)$, assuming that the random process
starts when the mode with wavenumber $k_{1}=e^{-1}a_{out}H_{0}$ crosses horizon. 
This procedure automatically subtracts the large-scale contribution
of modes with $k < k_{1}$. The small-scale contribution is also absent
due to our absorbing boundary condition. We have
$${\cal P}(h)={1\over \sqrt{2\pi \delta_{pl}^{2}}}{1\over{(x+1)}^{3/2}}
\exp{\lbrace-{x^{2}\over 2\delta_{pl}^{2}(x+1)}\rbrace} \eqno  39$$
from the eqn. $(35)$, where $x=\ln(1+h)$, and in the limit of small $h$ we
obtain again the Gaussian distribution
$${\cal P}(h)\approx {\cal P}_{G}(h)={1\over \sqrt{2\pi \delta_{pl}^{2}}}
\exp{\lbrace-{h^{2}\over 2\delta_{pl}^{2}}\rbrace}. \eqno 40$$
The distribution $(39)$ has nonzero first momentum 
$M_{1}=\int^{\infty}_{-1}dhh{\cal P}(h)={3\over 2}\delta_{pl}$ (the lower
limit of integration should be -1, since the metric perturbations with
$h < -1$ are cut off). The contribution of $M_{1}$ should be added to
the background part of metric, and further we will use the renormalized metric
perturbation $h_{r}=h-{3\over 2}\delta_{pl}$ instead of $h$.
The probability to find the metric perturbations $h_{r}$ 
with amplitude greater
than some threshold value 
$h_{*}$ ${\it P}(h_{*})=\int^{\infty}_{h_{*}}dh{\cal P}(h)$ can 
be estimated as 
$${\it P}(h_{*})\approx {1\over \sqrt{2\pi}}
({2\delta_{pl}{(x_{*}+1)}^{1/2}\over x_{*}(x_{*}+2)})
\exp{\lbrace-{x_{*}^{2}\over 2\delta_{pl}^{2}(x_{*}+1)}\rbrace}, \eqno 41$$
where $x_{*}=\ln(1+{3\over 2}\delta_{pl}+h_{*})$, 
and we assume $h_{*} \gg \delta_{pl}$.
The same quantity, but calculated for the Gaussian distribution takes the
well-known form
$${\it P}_{G}(h)\approx {1\over \sqrt{2\pi}} {\delta_{pl}\over h_{*}}
\exp{\lbrace-{h_{*}^{2}\over 2\delta_{pl}^{2}}\rbrace}. \eqno 42$$

The observed quantities (such as, e.g. the matter density of PBH's in different
cosmological epochs) can be easily expressed in terms of the probability ${\it P}(h_{*})$,
provided the mass of PBH's and some criterion for PBH's formation are fixed.
In our case the criterion for PBH's formation should give the information about
the threshold value $h_{*}$. Since this criterion plays very important role,
let us discuss it in some details. 
First let us note that PBH's are formed from high amplitude peaks in the
density distribution which are approximately spherically-symmetric (see e.g.
Ref. [34]).
It can also be easily shown that the maxima in the matter density
correspond to the maxima in the function $a_{ls}(\vec x)$. 
The form of $a_{ls}(\vec x)$ totally 
specifies the number of regions going to
PBH's as well as dynamics of the collapsing regions.
Therefore we formulate the criterion of PBH's formation
in terms of conditions imposed on the function $a_{ls}(\vec x)$.
 
The first criterion was formulated by Carr in his seminal paper [14].
It was shown that an over-dense region forms PBH
if the density contrast at the horizon scale ${\delta \rho \over \rho}$
lies approximately within the limits 
${1\over 3} < {\delta \rho \over \rho} < 1$.
The first part of this inequality tells that the over-dense region should stop
expansion before the scale of the region crosses the sound horizon.
The second part requires that the over-dense region 
does not collapse before crossing the causal horizon, and consequently
the perturbation does not produce a closed world separated from
the rest of the Universe.
Then the criterion for PBH's formation was improved 
by Nadegin, Novikov and Polnarev [21] (hereafter NNP), 
and also by Biknell and Henriksen
[22] with help of numerical computations.
The initial condition used by NNP was chosen as a non-linear
metric perturbation having the form of a part of the closed Friedman Universe
matched with the spatially flat Universe through an intermediate layer of 
negative density perturbation. The conditions for PBH's
formation depend on the size of this part 
(i.e. the amplitude of the perturbation), as well as on the size of the 
matching layer. 
The smaller matching layer is, the larger the pressure gradients needed to
prevent collapse will be.
Therefore, the amplitude of the
perturbation forming PBH  must be greater in  the case of narrow intermediate layer. 
In terms of our function $a(\vec x)$ the NNP criterion reads 
$$h_{*}\equiv {a_{+}\over a_{-}}-1 > 0.75-0.9 \eqno 43$$
where $a_{+}$ is the value of $a(\vec x)$ at the maximum of the perturbation
and $a_{-}$ is the same quantity outside the perturbed region
\footnote{In the linear theory 
the density perturbation at horizon scale relates to the metric perturbation
by ${\delta \rho \over \rho}={4\over 9} h$ (see, e.g. [27]).
Therefore the estimate $(43)$ is in agreement with Carr's result.}.
The first number on the right hand side of $(43)$ corresponds to the matching
layer of size comparable with size of the over-dense region, and the second
number corresponds to the narrow matching layer. Assuming
the matching layer to be sufficiently large we take $h_{*}=0.75$ as a
criterion of PBH's formation.

Once the criterion is specified, we can link the desired PBH's abundance
$\beta (M_{pbh})\approx P(h_{*pbh})$ with the parameters of our model. 
For instance, consider the model having the matter density of
PBH's equal to the critical one (the density parameter $\Omega_{pbh}=1$). 
In this model we have [3], [6]
$$\beta (M)=10^{-8}{({M\over M_{\odot}})}^{1/2}. \eqno 44$$
Equating the expression $(44)$ to the probability function
$(39)$, we have the equation determining the  
amplitude $\delta_{1pl}$ required for PBH's abundance $(44)$ as a function
of $M_{pbh}$
$${\it P}(h_{*pbh}, \delta_{1pl})=\beta(M_{pbh}), \eqno 45$$
and equating the expressions $(42)$ and $(44)$ we obtain the analogous
equation for determining the reference amplitude $\delta_{2pl}$ when
the non-Gaussian effects are switched off. 
The solution 
of these equations is given in Fig. 1. 
\begin{figure}[h]
\vspace{0.3cm}\hspace{-1cm}\epsfxsize=9cm
\epsfbox{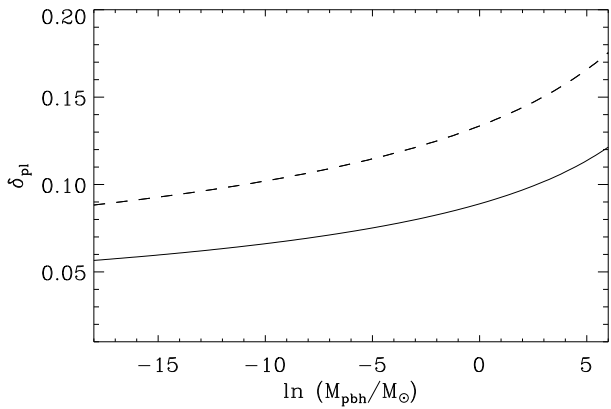}
\caption{We plot the dependence of plateau parameter $\delta_{pl}$ on PBH's 
mass $M_{pbh}$ assuming that the PBH's abundance is given by the eqn. $(44)$. The solid
line represents the solution of eqn. $(45)$ (i.e we calculate $\delta_{pl}$
taking into account the non-Gaussian effects in this case). The dashed line
represents $\delta_{pl}$ calculated in the standard Gaussian theory.
The PBH's masses lie in the range: $10^{-18}M_{\odot} < M_{pbh} < 10^{6}M_{pbh}$.
The PBH's of the mass $10^{-18}M_{\odot}\sim 10^{15}g$  should be evaporated
at the present time. Actually, the abundance of these PBH's is limited much
stronger than is assumed in our calculations.}
\end{figure} 
One can see from this Fig. 
that the quantities 
$\delta_{1pl}$ and $\delta_{2pl}$ increase 
with increasing of $M_{pbh}$ and $\delta_{1pl}$ is always smaller than 
$\delta_{2pl}$. It means that non-Gaussian effects over-produce PBH's
in our model (at least when the simple criterion $(43)$ is used), and
the slope of potential can be steeper than that required in the Gaussian
case. Typically, the ratio ${\delta_{2pl}\over \delta_{1pl}}$ is about
$1.5$. Say, for the case of $M_{pbh}=M_{\odot}$, we have 
$\delta_{1pl}(M_{\odot})\approx 0.089$ and
$\delta_{2pl}(M_{\odot})\approx 0.134$. 
We plot the probability function
${\cal P}(h)$ for $\delta_{1pl}(M_{\odot})=0.089$ in Fig. 2. 
\begin{figure}[h]
\vspace{0.3cm}\hspace{-1cm}\epsfxsize=9cm
\epsfbox{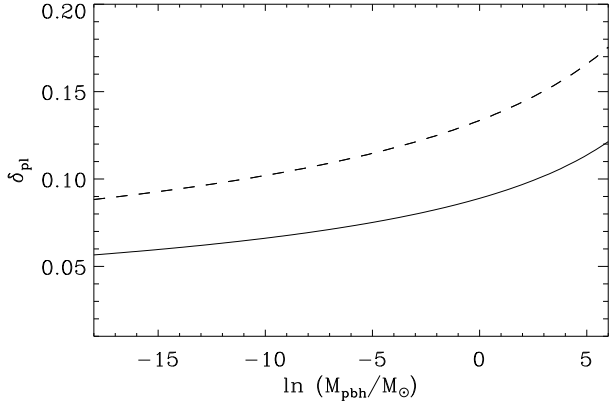}
\caption{The dependence of probability density ${\cal P}(h)$ on the metric amplitude
$h$. The non-Gaussian curve (solid line) is calculated with help of eqn. $(39)$ assuming
PBH's abundance $\beta(M_{\odot})\approx 10^{-8}$. That gives 
$\delta_{1pl}(M_{\odot})\approx 0.089$. The dashed line is the reference
Gaussian probability density calculated for the same abundance.
For that curve we have  $\delta_{2pl}(M_{\odot})\approx 0.134$. The dotted curve 
represents the Gaussian distribution taken with $\delta_{1pl}(M_{\odot})\approx 0.089$. 
This distribution strongly under-produces PBH's, and in this case we have $\beta \sim
10^{-17}$.} 
\end{figure} 

In this Fig., we also
plot the Gaussian probability function ${\cal P}_{G}(h)$ for 
$\delta_{2pl}(M_{\odot})=0.134$ (dashed line) and the same quantity
for $\delta_{1pl}(M_{\odot})=0.089$ (dotted line). 
Comparing the curves
that correspond to the same PBH's abundance, we see that the non-Gaussian curve
is flatter having larger values of ${\cal P}(h)$ 
at large $h$. The values of the Gaussian
curve with the same plateau parameter $\delta_{1pl}(M_{\odot})$ 
is smaller
by many orders of magnitude than the values of the non-Gaussian curve
in the case of large $h$.

Finally, let us note, that the non-Gaussian effects does not
modify significantly the estimates based on the Gaussian theory. As we have
seen, the ambiguity in the choice of the plateau slope due to these effects is
about $1.5$. This ambiguity seems to be less than the ambiguity in
other parameters and can be obviously absorbed by a small 
change of the potential slope.  

\section{Discussion}

We have demonstrated that the non-Gaussian effects related to the dynamics of the 
coarse-grained field (inflaton) and to the evolution of the large-scale part of metric 
over-produce
large-amplitude inhomogeneities of the metric compared to the prediction of the
Gaussian (linear) theory of perturbations. We have derived an analytical expression 
for non-Gaussian 
probability distribution for non-linear metric perturbations, and estimated the influence of
non-linear effects on the probability of primordial black holes formation. We used the simple
single field inflationary model with peculiarity in form of the flat region in inflaton
potential $V(\phi)$, and power-law slope of the potential outside the peculiarity region.
The key point of our approach is in the using  of inhomogeneous coarse-grained metric
function $a(\vec x)$ instead of the coarse-grained field $\phi_{ls}$ as a basic
quantity. This allows to match the physical condition of production of inhomogeneities  
during inflation with the "observable" quantities.
 
Our results can be considered as semi-qualitative only. The uncertainties come
fromthe phenomenological character of our inflationary model as well as from 
the oversimplified treatment of the process of PBH's formation. The uncertainties related 
to the choice of parameters of inflationary model are mainly due to unknown form of the
potential between the steep and flat regions, and also due to our friction-dominated 
assumption in the consideration of the stochastic process. These uncertainties can be   
eliminatedwith help of numerical simulations of stochastic process in more realistic 
models of inflation.
The ambiguities concerning the criterion of PBH's formation are mainly due to the 
one-point treatment of this process. Actually, PBH formation is nonlocal,
and dynamics of collapsing region depends strongly on the form of the spatial profile of
the density perturbation (see e.g. Refs. [22], [35] for discussion of this point).
The form of the spatial profile can be studied by means of n-point correlation functions
of the coarse-grained metric and field. Unfortunately, the formalism of n-point 
correlation functions is still not elaborated (see, however the Ref. [35] for the
first discussion). Note, that probably the influence of the spatial profile of the collapsing
region may be taken into account by a redefinition of the threshold value $h_{*}$, and
this value might be effectively less. In this case the role of the non-linear effects would
be damped.

Finally we would like to note that the form of the distribution $(35)$
does not depend explicitly on the specific parameters of our model. This allows to suppose
that similar distributions can be obtained in more complicated models, say, in
two-field models proposed in Refs. [8], [9]. We are going to check this very
interesting assumption in our future work.

\bigskip
{\large\bf Acknowledgments}
\bigskip

The author acknowledges P. Naselsky and I. Novikov 
for many valuable discussions, and also A. Beloborodov, A. Dolgov and
D. Markovic for useful comments. This work was supported in part by the Danish 
Research Foundation through its establishment of the Theoretical Astrophysic

\end{document}